# Seeing Earth's Orbit in the Stars: Parallax and Aberration

*Todd Timberlake, Berry College, Mount Berry, GA*

---

During the 17th century the idea of an orbiting and rotating Earth became increasingly popular, but opponents of this view continued to point out that the theory had observable consequences that had never, in fact, been observed.[1] Why, for instance, had astronomers failed to detect the annual parallax of the stars that *must* occur if Earth orbits the Sun?

To address this problem, astronomers of the 17th and 18th centuries sought to measure the annual parallax of stars using telescopes. None of them succeeded. Annual stellar parallax was not successfully measured until 1838, when Friedrich Bessel detected the parallax of the star 61 Cygni.[2] But the early failures to detect annual stellar parallax led to the discovery of a new (and entirely unexpected) phenomenon: the aberration of starlight. This paper recounts the story of the discovery of stellar aberration. It is accompanied by a set of activities and computer simulations that allow students to explore this fascinating historical episode and learn important lessons about the nature of science.[3]

**Parallax**

Hold up your thumb in front of your face at arm's length. Close your right eye and view your thumb against the background. Now open your right eye and close your left: you will see your thumb appear to move against the background. The apparent displacement of your thumb is really caused by the displacement of your observing location (from your left eye to your right). This phenomenon, known as parallax, plays an important role in astronomy.

The Ancient Greeks were aware that celestial objects viewed from different locations on Earth might appear in slightly different locations relative to the background stars. They used this effect, known as *diurnal parallax*, to accurately estimate the distance to the Moon. Figure 1 illustrates the geometry of this measurement. For simplicity we will assume the Moon lies in the equatorial plane and observations are made from opposite sides of Earth's equator (the points $O_1$ and $O_2$). The two observers see the Moon at different locations on the Celestial Sphere (or the starry background, to use more modern terminology). These apparent locations are separated by an angle $2\theta$. From trigonometry (and using the small angle approximation) we find:

$$\theta = \frac{90b}{\pi d}, \tag{1}$$

where θ is the parallax angle in degrees, *b* is the baseline (in this case the diameter of Earth), and *d* is the distance from the center of Earth to the center of the Moon.

Note that if you can measure the parallax angle, and you know the baseline, you can determine the distance to the object. This is how the Ancient Greeks determined the distance to the Moon. But they could not detect any parallax in the stars. This was easily explained: the stars were very far away compared to the Earth's diameter ($b<<d$), so the parallax angle was too small to measure.

This situation changed with the introduction of the Copernican system. Copernicus was well aware of parallax. In fact, parallax effects explain the motion of the Sun and certain aspects of planetary motions in the Copernican system. But if Earth has a yearly orbit around the Sun, then the stars should also display a parallactic wobble with a period of one year. (The `AstronomicalParallax2D` program illustrates this *annual parallax* for an object in the Earth's orbital plane, as well as the diurnal parallax of an object in the equatorial plane.[3]) No such effect was seen. Copernicus relied on the same explanation used by the Ancient Greeks: the stars are just too far away. But Copernicus had to claim that the stars are very far away compared to the *diameter of Earth's orbit*, which meant that the stars were vastly farther away than the Ancient Greeks believed.

The lack of detectable annual stellar parallax made some astronomers skeptical of the Copernican system. The stage was set for a race to measure the annual parallax of a star. Measuring this parallax would not only vindicate the Copernican system, but also provide a direct measurement of the distance to the star.

**Searching for Annual Parallax**

In 1669 the English scientist Robert Hooke attempted to measure the annual parallax of the star gamma Draconis. Hooke chose gamma Draconis because it passes nearly overhead in London, so his observations would not be significantly affected by atmospheric refraction. To carry out his measurement, Hooke built a zenith telescope into his Gresham College apartments. To use this telescope Hooke had to lay down below the eyepiece on the ground floor and look up through holes in the upper floor and roof, and finally through an objective lens in a tube that jutted from the top of the building. With this zenith telescope Hooke measured the angle between the zenith point (straight up) and gamma Draconis when that star crossed the meridian (the north-south line in the sky).[4]

Annual parallax would show up as a periodic variation in this zenith distance with a period of one year. Since gamma Draconis does not lie within Earth's orbital plane (in fact, it lies nearly perpendicular to that plane), we must consider a full three-dimensional picture of the situation rather than the simplified 2D situation of Fig. 1. The `AstronomicalParallax3D` program illustrates the three-dimensional geometry for a star in any direction on the sky.[3] This program was used to construct a plot of the Declination of Gamma Draconis as a function of time, as shown in Figure 2. Declination measures the north-south position of a star in the sky and thus corresponds to Hooke's zenith angle measurements. In Fig. 2 the amplitude has been greatly exaggerated in order to make the effect visible in the simulation, but the plot illustrates the correct pattern of changes in Declination resulting from annual parallax.

In 1674 Hooke published his results and claimed to have detected annual parallax.[5] Although he gave a detailed discussion of his careful measurement procedure, he presented only four observations. Problems with the telescope and his own health prevented him from continuing the work. Hooke's data is shown in Figure 3. A comparison with the parallax prediction in Fig. 2 shows that Hooke's data seems to match the expected pattern for annual parallax, but his contemporaries did not find four observations made with an unreliable telescope very convincing. The race for parallax was not yet over.

**Aberration of Starlight**

Several astronomers attempted to follow up on Hooke's measurement. G. D. Cassini and Jean Picard measured variations in the position of Polaris, but with inconclusive results. Astronomer Royal John Flamsteed thought he had measured an annual parallax for Polaris until Cassini pointed out that Flamsteed's data did not fit the expected pattern.[6]

In 1725, Samuel Molyneux and James Bradley set out to repeat Hooke's measurements of Gamma Draconis by constructing a zenith telescope in Molyneux's mansion at Kew near London. They found that the star varied its position, but not in the way reported by Hooke. Bradley followed up on this work by measuring several more stars using a shorter zenith telescope in his residence at Wanstead. Fig. 3 shows Bradley's data for Gamma Draconis and Alkaid (Eta Ursa Majoris). The data show the sinusoidal variation expected for parallax, but the phase is 3 months off from the predictions shown in Fig. 2!

Bradley knew of Ole Rømer's 1676 estimate of the finite speed of light, and eventually used this idea to devise an explanation for his data. Bradley realized that it takes the light from a star a finite time to travel through a telescope tube, and during this time the tube moves slightly because of the motion of Earth. Bradley's theory is illustrated in Figure 4 for the case of a star that lies directly overhead. The starlight enters the top of the tube at $\mathbf{p_1}$. It then travels a distance $h$ before reaching the eyepiece at $\mathbf{p_3}$. If $c$ is the speed of light, then it takes a time $h/c$ for the light to travel this distance. During this time the telescope has moved a distance $vh/c$, where $v$ is the velocity of Earth's orbital motion.[7] As a result, the telescope cannot be aimed directly up toward the star. It must be tilted slightly toward the direction of the telescope's motion. This phenomenon is now called the aberration of starlight. Applying trigonometry to Fig. 4 we find that

$$\tan \theta = \frac{v}{c}, \tag{2}$$

where θ is the angle of tilt. For an interactive animated version of Fig. 4 see the `StellarAberration2D` program.[3]

Fig. 4 illustrates the apparent displacement of the star due to aberration at a single moment in time. As the Earth moves around in its orbit, its velocity changes direction and therefore the displacement of stars due to aberration will change.[8] The pattern of apparent movement depends on the location of the star in the sky. These patterns are illustrated in the `StellarAberration3D` program.[3] This program was used to construct plots of the apparent declination of Gamma Draconis and Alkaid, using Bradley's theory of stellar aberration, as shown in Figure 5. A comparison with Fig. 4 shows that the pattern predicted by Bradley's theory fits his observational data.

Bradley's data indicates a displacement of 20.2 seconds of arc for a star that lies in a direction perpendicular to Earth's motion. Using Equation 2 Bradley found that the speed of light must be 10,210 times as great as Earth's orbital speed, so light takes 8 minutes, 12 seconds to travel from the Sun to Earth.[9]

Bradley's theory explained the movement of Polaris observed by Cassini, Picard, and Flamsteed, but he could not give an explanation for Hooke's measurements of Gamma Draconis. Hooke apparently saw what he wanted to see amidst a storm of instrumental error. Bradley showed that his data fit the aberration theory so well that any difference, which might be due to annual parallax, was probably less than one half second of arc. This meant that Gamma Draconis was more than 400,000 times farther from us than the Sun.

It was impossible to measure parallax using Hooke's method without first knowing about aberration. As Eq. 2 shows, the angular displacement due to aberration does not depend on the distance to a star. But as Eq. 1 shows, the parallax angle decreases with distance. Since Gamma Draconis is relatively bright, and thus probably nearby, it might be expected to have one of the larger parallaxes. Even so, Bradley had shown that aberration completely swamped parallax for Gamma Draconis, and for more distant stars the situation would only get worse. It was only by measuring the relative motions of two stars, close together in our sky but far apart in space, that astronomers like Bessel, Struve, and Henderson would finally measure the annual parallax of stars in the 1830s. This double star method, however, would not have led to the discovery of aberration because aberration affects both stars equally.[10]

**Lessons**

There are several lessons that can be learned from the stories of Hooke and Bradley.[11] Hooke's claim to have measured parallax shows us that reliable conclusions cannot be drawn from a small number of measurements. A conclusive demonstration requires a series of many measurements, all of which fit a definite pattern like the one established by Bradley. Bradley's story also shows us that scientists sometimes succeed in unexpected ways. Bradley sought to confirm Earth's orbital motion by detecting annual parallax. Instead, he confirmed it by detecting and explaining stellar aberration. But his success was not complete: a parallax measurement would have given the distance to the star, while a measurement of aberration provides no information about stellar distances.

Bradley's theory of aberration also shows us that good scientific explanations can draw together untested hypotheses to explain something in a way that gives support to all of the hypotheses. Neither the orbit of Earth nor the finite speed of light had been independently confirmed before 1725 (although Newton's *Principia* had left little doubt about Earth's orbit), but Bradley showed that his data were easily explained if both of these hypotheses were true.

Finally, these stories show that assigning credit for a "discovery" is tricky business. Hooke claimed to measure parallax, but his claim is now discredited. Flamsteed measured aberration but thought he was measuring parallax. Cassini recognized that Flamsteed was measuring something other than parallax, but it was Bradley who *explained* his data and therefore is credited with the discovery of aberration.

The activities and computer simulations in Ref. 3, which are designed to guide students through an exploration of this historical episode, illustrate these important lessons about the nature of science.

**Figures**

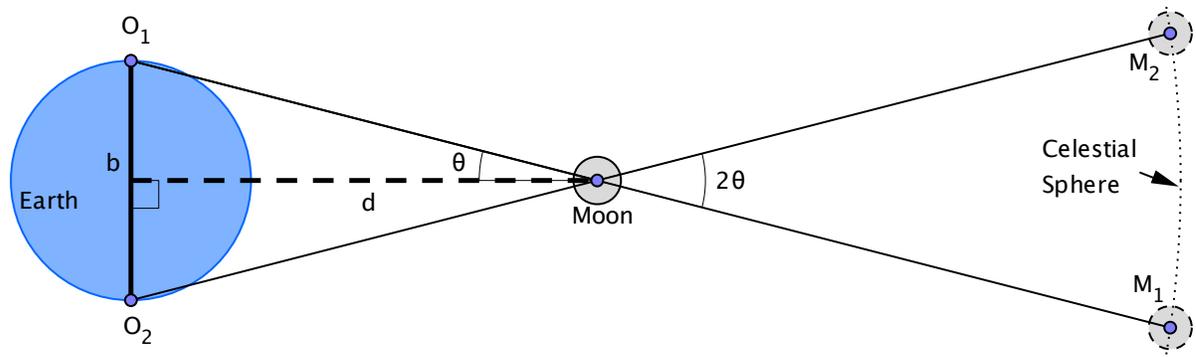

Figure 1: Parallax of the moon, viewed from opposite sides of Earth. NOT TO SCALE!

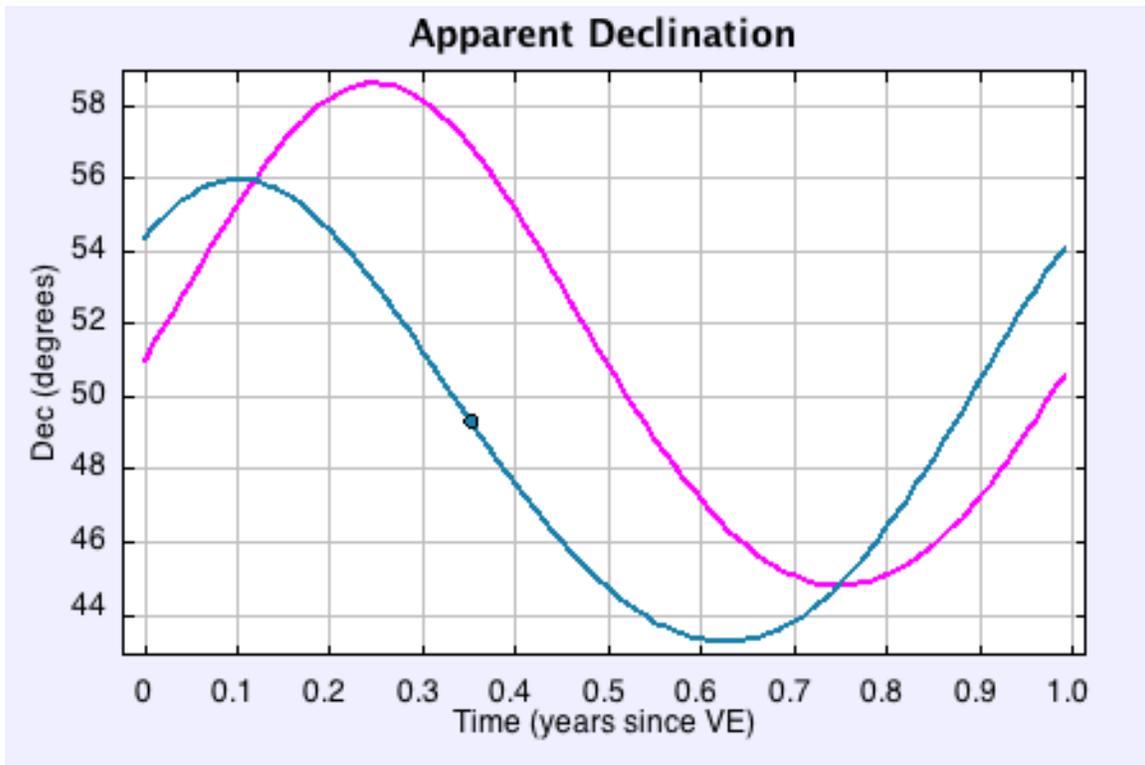

Figure 2: Predicted variation in Declination resulting from the annual parallax of Gamma Draconis (magenta) and Alkaid (blue). The amplitude of the variation is greatly exaggerated so as to be visible in the simulation.

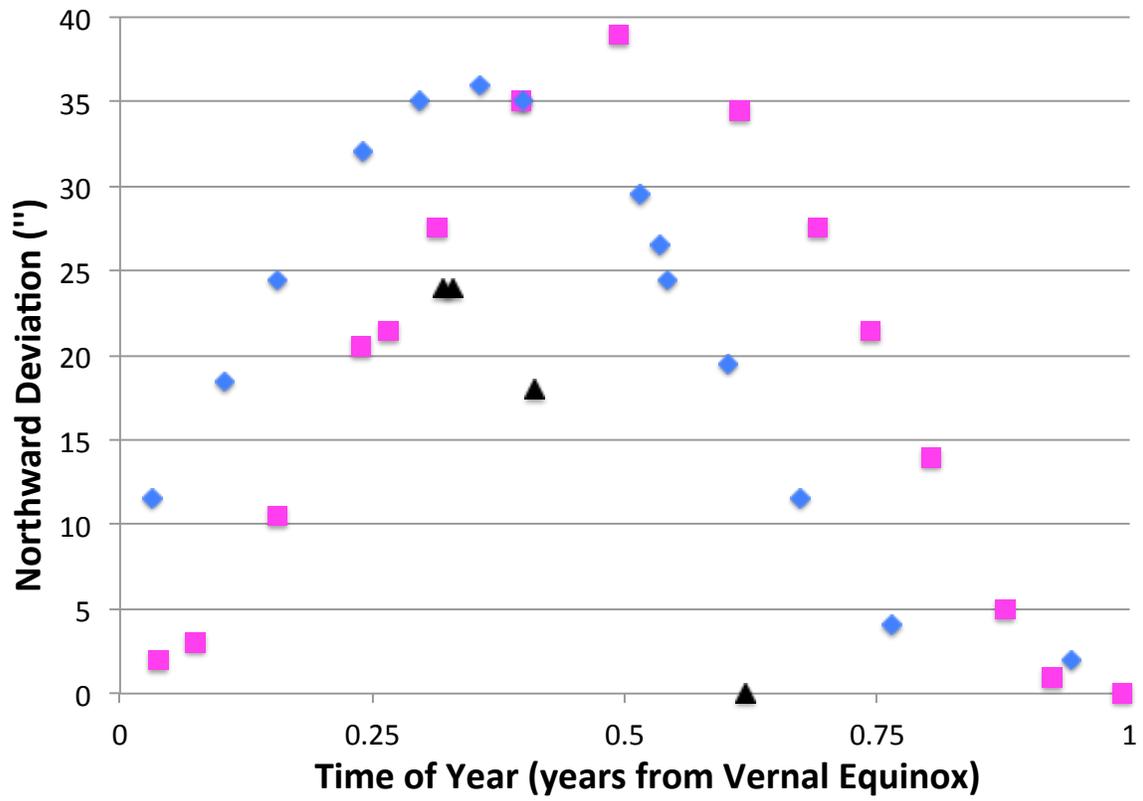

Figure 3: Observed northward deviation of stars. The data points show Hooke's data for Gamma Draconis (black triangles) and Bradley's data for Gamma Draconis (magenta squares) and Alkaid (blue triangles).

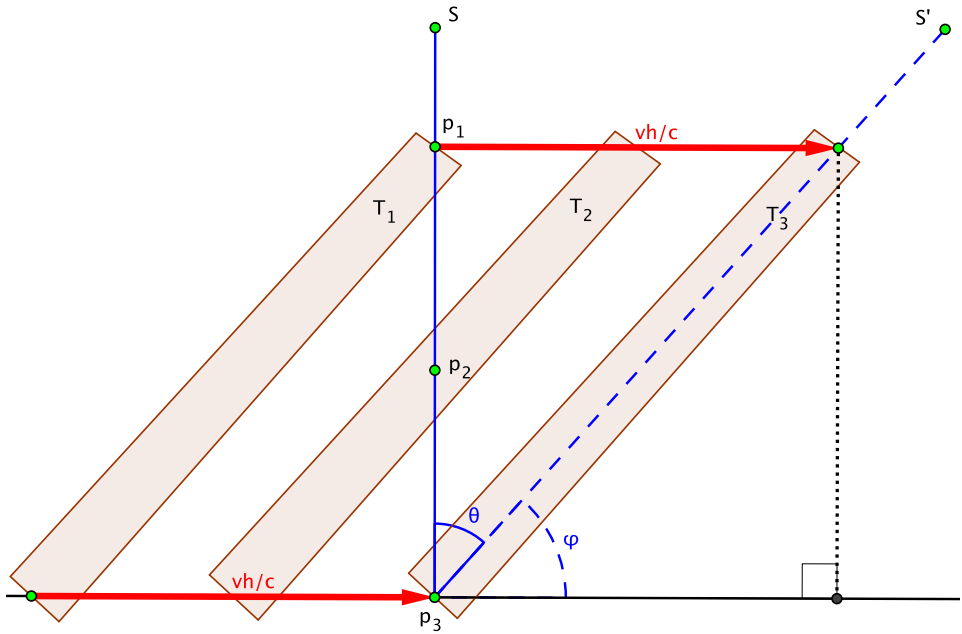

Figure 4: Illustration of the aberration effect, showing the telescope tube T at three different times. Because the telescope moves a distance *vh/c* during the time it takes light to travel down the tube, the tube must be tilted in the direction of motion in order for the light to move along the optical axis of the tube. So a star at S will appear to be at S'.

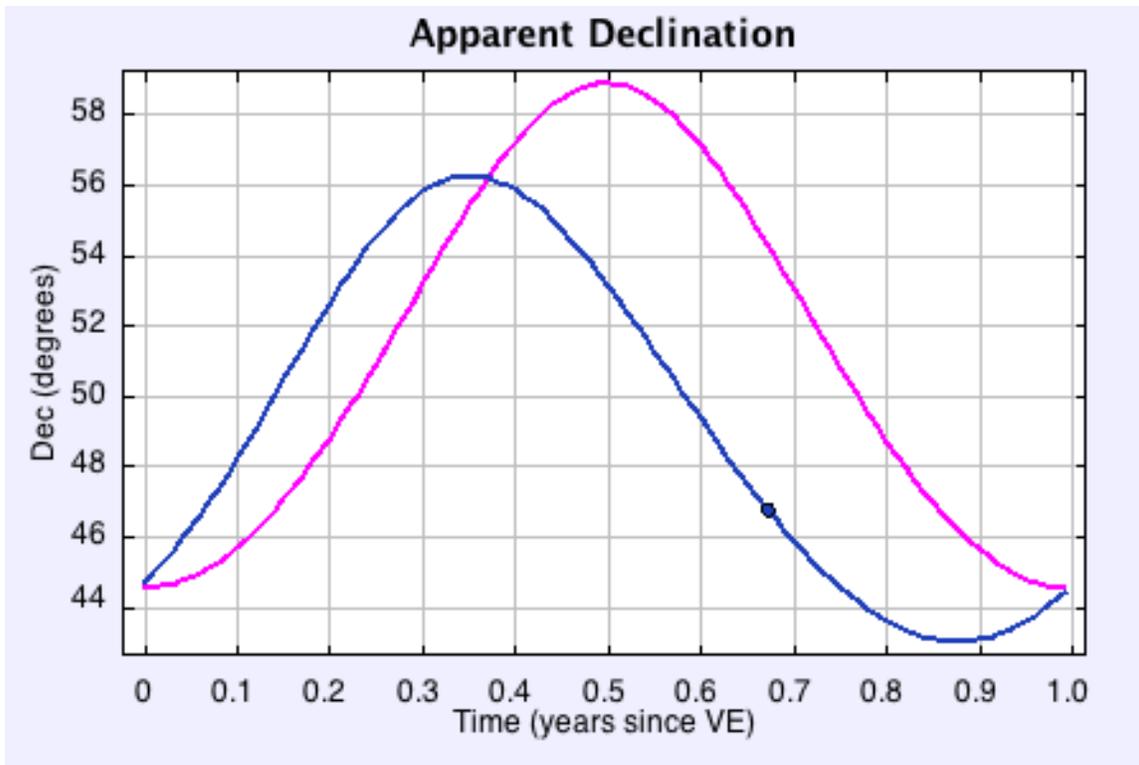

Figure 5: Predicted variation in Declination resulting from aberration for Gamma Draconis (magenta) and Alkaid (blue). The amplitude of the variation is greatly exaggerated so as to be visible in the simulation.

---

[1] Christopher M. Graney, "Teaching Galileo? Get to Know Riccioli! What a Forgotten Italian Astronomer Can Teach Students About How Science Works," *Phys. Teach.* **50**, 18-21 (2012).
[2] Alan Hirshfeld, *Parallax: The Race to Measure the Cosmos* (W H Freeman, 2001).
[3] A self-contained Java program containing all of the simulations is available at www.opensourcephysics.org/items/detail.cfm?ID=12029. The activity handouts are available as supplementary documents on the same page.
[4] Hooke observed gamma Draconis in daylight, as well as at night. He may have been the first person to observe a star through a telescope during the day.
[5] Robert Hooke, *An Attempt to Prove the Motions of Earth by Observations* (London, 1674). Available online at http://www.roberthooke.com/motion_of_the_earth_001.htm.
[6] M. E. W. Williams, "Flamsteed's Alleged Measurement of Annual Parallax for the Pole Star," *J. Hist. Astron.* **10**, 102-116 (1979).
[7] The telescope also moves due to Earth's rotation, but the rotational velocity is about 64 times smaller than the orbital velocity, so to a good approximation we can ignore the effect of rotation. In any case the effect Bradley was trying to explain had an *annual* period.

[8] Note that a fixed aberration angle would be undetectable. For a discussion of this point, and how aberration relates to special relativity, see Thomas E. Phipps, Jr. "Relativity and aberration," *Am. J. Phys.* **57**, 549-551 (1989).

[9] James Bradley, "A Letter to Dr. Edmond Halley Astronom. Reg. &c. giving an Account of a new discovered Motion of the Fix'd Stars," *Phil. Trans.* **35**, 637-661 (1729).

[10] I'd like to thank Charlie Holbrow of Colgate University for this insight.

[11] Michael Hoskin, *Stellar Astronomy: Historical Studies* (Science History Publications, 1986), pp. 29-36.